\documentclass[12pt,a4paper]{article} 

\begin{document}

\title{All ``static" spherically symmetric perfect fluid solutions of Einstein's equations with constant equation of state parameter and finite-polynomial ``mass function"}

\author{{\.{I}brahim Semiz\thanks{mail: ibrahim.semiz@boun.edu.tr}}\\    
   Bo\u{g}azi\c{c}i University, Department of Physics\\
Bebek, \.{I}stanbul, TURKEY\\}
\date{ }

\maketitle

\begin{abstract}
We look for ``static" spherically symmetric solutions of Einstein's Equations for perfect fluid source with equation of state $p=w\rho$. In order to include the possibilities of recently popularized dark energy and phantom energy possibly pervading the spacetime, we put no constraints on the constant $w$. We consider all four cases compatible with the standard ansatz for the line element, discussed in previous work. For each case we derive the equation obeyed by the mass function or its analogs.
For these equations, we find {\em all} finite-polynomial solutions, including possible negative powers. 

For the standard case, we find no significantly new solutions, but show that one solution is a static phantom solution, another a black hole-like solution. For the dynamic and/or tachyonic cases we find, among others, dynamic and static tachyonic solutions, a Kantowski-Sachs (KS) class phantom solution, another KS-class solution for dark energy, and a second black hole-like solution.

The black hole-like solutions feature segregated normal and tachyonic matter, consistent with the assertion of previous work. In the first black hole-like solution, tachyonic matter is inside the horizon, in the second, outside.

The static phantom solution, a limit of an old one, is surprising at first, since phantom energy is usually associated with super-exponential expansion. The KS-phantom solution stands out since its ``mass function" is a ninth order polynomial.
\end{abstract}

\section{Introduction and Motivation}

Exact solutions of Einstein's Field Equations
\begin{equation}
G_{\mu\nu} = \kappa T_{\mu\nu}               \label{ee}
\end{equation}
are, of course, of interest for various purposes (Here,  $G_{\mu\nu}$ is the Einstein tensor, $T_{\mu\nu}$ the stress-energy-momentum tensor and $\kappa$ the coupling constant). Since the equations are very complicated, to find solutions one often makes simplifying assumptions about the left-hand-side and/or the right-hand-side. Popular simplifying assumptions about the left-hand-side include staticity and spherical symmetry. As is well known, the use of both assumptions together leads to the ansatz~\cite[Sect.23.2]{mtw}
\begin{equation}
ds^{2} = -B(r) dt^{2} + A(r) dr^{2} + r^{2} d\Omega^{2}   \label{ansatz}
\end{equation}
for the metric.

Most-often used simplifying assumptions about the right-hand-side of (\ref{ee}) are that $T_{\mu\nu}$ represents vacuum (i.e. vanishes) or an electromagnetic field or a perfect fluid. For example, the vacuum assumption, together with the ansatz (\ref{ansatz}) gives uniquely the Schwarzschild metric, the simplest and best-known black hole solution.

The perfect fluid form of $T_{\mu\nu}$ is
\begin{equation}
T_{\mu\nu} = (\rho + p) u_{\mu} u_{\nu} + p g_{\mu\nu}  \label{pfemt}
\end{equation}
where $\rho$ and $p$ are the energy density and pressure, respectively, as measured by an observer moving with the fluid, and $u_{\mu}$ is its four-velocity. The use of this $T_{\mu\nu}$ together with ansatz (\ref{ansatz}) describes the interiors of static spherically symmetric stars, for example. But the description (\ref{pfemt}) is not complete: $\rho$ and $p$ should also be specified as functions of particle number density, temperature, etc. One further simplifying assumption, justified under most circumstances, is that there is a relation, called an {\em equation of state} $f(p,\rho)=0$ between $p$ and $\rho$. In cosmology, one usually assumes that the equation of state is a proportionality,
\begin{equation}
p=w\rho,     \label{eos}
\end{equation}
with e.g. $w=0$ describing the matter-dominated (or "pressureless dust") case, \mbox{$w=1/3$} the radiation-dominated case, \mbox{$w<-1/3$} dark energy, and \mbox{$w<-1$} phantom energy. The latter two concepts have been introduced into cosmology in the last decade  \cite{de,phantom}, after the discovery of the acceleration of the expansion of the universe \cite{acceleration-hiZsst,acceleration-SCP}.

Now that a good case exists that the universe might be dominated by dark energy, even phantom energy, one should look for exact solutions with these sources. In particular, static spherically symmetric solutions would be the easiest to find and might be relevant in the contexts of black holes or static stars. These solutions can be found starting from the ansatz (\ref{ansatz}), which for ``static" perfect fluid source, (i.e. $u^{\mu} = u^{0} \delta_{0}^{\mu}$) leads to the well-known Oppenheimer-Volkoff (OV) equation~\cite{ov}
\begin{equation}
p'   =  - \frac{(\kappa p r^{3} + F)}{2 r (r-F)} (\rho + p)     \label{ov}
\end{equation}
where
\begin{equation}
F(r) = \kappa \int \rho r^{2} dr          \label{FDefNS}
\end{equation}
is written as $F$ for brevity, and prime denotes $r$-derivative. $F(r)$ can be recognized as $\kappa/4\pi$ times the "mass function" defined in the literature. Into the OV equation (\ref{ov}) one must put $p$ in terms of $\rho$ via the equation of state, then $\rho$ in terms of $F'$, via (\ref{FDefNS}), eventually getting a differential equation for $F$. After solving for $F$, the metric functions can be found via 
\begin{eqnarray}
A(r) & = & \frac{r}{r-F(r)}     \label{EEns4} \\ 
\frac{B'(r)}{B(r)}   & =  & \frac{\kappa p r^{2} + 1}{r-F(r)}  -   \frac{1}{r} .  \label{EEns5}
\end{eqnarray}
The solutions can be interpreted as static only for positive $A(r)$ and $B(r)$, however.  In general,  the ansatz (\ref{ansatz}) admits four classes of solutions, called NS (the standard case), TD, ND (corresponding to Kantowski-Sachs~\cite[Sect.15.6.5]{exsols}, \cite{ks} case) and TS in~\cite{revisited}. The ND and TD solutions are not static, hence the quotes on ``static" in the title and abstract. For each class, one gets a different OV-like equation.

The OV equation is valid in case NS. For equation of state (\ref{eos}), it becomes
\begin{equation}
(w+1) F' (wrF'+F) + 2 w (rF''-2F')(r-F) = 0                      \label{MFeq-NS}
\end{equation}
where we put no constraint on $w$ other than that it is a constant.  This is a nonlinear equation whose general solution is difficult to find, unless $w=-1$ or $w=0$, so that one half or the other of (\ref{MFeq-NS}) vanishes. For other $w$, one can attempt a series solution
\begin{equation}
F(r)=\sum_{n=0}^{\infty} a_{n} r^{n}		\label{series}
\end{equation}
but the recursion expression one gets for $a_{n}$ involves all of $a_{0} \dots a_{n-1}$ and it seems not possible to even show that (\ref{series}) converges, let alone find a closed expression for $a_{n}$.

We can, however, find {\em all} of the {\em finite-polynomial} solutions of (\ref{MFeq-NS}). This we do in the next section. In fact, we find all finite {\em Laurent} polynomials, i.e. we consider also negative powers\footnote{In the rest of this work,  we will use ``power'' also when we really mean ``order of the power''. It should be clear from the context which meaning is intended.} of $r$, but find none in case NS.  Four  of the found solutions are valid for particular values of $w$, and two for general $w$. While none of the solutions is totally original, the procedure shows that there are no other finite-polynomial solutions; and in Section \ref{sect:NSdisc} we discuss properties of the spacetimes. 

In Section \ref{FindOthSols}, we find all finite-polynomial solutions for $F(r)$ in the TD, ND and TS cases;
and in Section \ref{sect:NonStdSols} we discuss their properties. In the short Section \ref{sect:PolA}, we also ask if we can find any solutions with finite-polynomial $A(r)$. Finally we conclude by pointing out the more interesting, and possibly original solutions; and emphasizing the main point of this work, that there are no other solutions under our restrictions.

\section{All finite-polynomial solutions for the mass function from the standard OV equation\label{FindStdSols}}

In case NS, any power of $r$ less than 3 in $F(r)$ means a diverging density at the origin; in particular, a constant term corresponds to a point mass there, while negative powers mean diverging mass function, and therefore seem unnatural. On the other hand, the meaning of $F(r)$ is different in the TD, ND(KS) and TS cases, therefore negative powers could be more acceptable.

Before starting the general case, however, let us first consider the special cases mentioned above, after eq.(\ref{MFeq-NS}); especially since they also give polynomial solutions. They are
\begin{equation}
{\rm {\bf Solution \; 1:}} \;\;\;\; w=-1, \;\; F(r) = A r^3  +  C      \label{Sol1}
\end{equation}
\begin{equation}
{\rm {\bf Solution \; 2:}} \;\;\;\; w=-1, \;\; F(r) = r      \label{Sol2}
\end{equation}
and
\begin{equation}
{\rm {\bf Solution \; 3:}} \;\;\;\; w=0, \;\; F(r) =  C.      \label{Sol3}
\end{equation}
where in Solutions 1 and 3, $C$ is a constant. 

Now, we will consider general (but constant) $w$. In some cases to be discussed below, the number of terms in the polynomial will be known. Then, substitution into eq.(\ref{MFeq-NS}) gives a certain number of terms, and the analysis reduces to straightforward, if possibly tedious, algebra, which can be expedited by the use of symbolic computation software. We will call this the "brute force" approach.    

  Otherwise, we will call the highest and lowest power of $r$ in $F(r)$, $m$ and  $\tilde{m}$. The second-highest, third-highest, second-lowest and third-lowest powers of $r$ in $F(r)$ we will call $n$, $p$, $\tilde{n}$ and $\tilde{p}$ respectively, when they exist; and $A$, $B$, $C$, $\tilde{A}$, $\tilde{B}$, $\tilde{C}$ will be the respective coefficients. Therefore
\begin{equation}
F(r) = A r^{m} + B r^{n} + C r^{p} + \ldots + \tilde{C} r^{\tilde{p}} + \tilde{B} r^{\tilde{n}} + \tilde{A} r^{\tilde{m}} 		\label{FpowersDef}
\end{equation}
if $F(r)$ has more than five terms. We will substitute the polynomial into the left-hand-side of  (\ref{MFeq-NS}) and set coefficients of all powers of $r$ equal to zero.

We start the general analysis by considering the coefficient of the highest power of $r$ in eq.(\ref{MFeq-NS}), for $m > 1$:
\begin{equation}
(w+1)m (wm+1) A^{2} - 2wm(m-3) A^{2} = 0.      \label{2m-1_coeff}
\end{equation}
Therefore in this case $A$ is arbitrary and
\begin{equation}
m = \frac{7w+1}{w(1-w)}  \;\;\;\;\;\;\; {\rm [for} \; m>1 ].         \label{m_from_w}
\end{equation}

If (\ref{m_from_w}) had given an integer, we would have found the order of the polynomial for arbitrary $w$. Since it does not, we conclude that in this case finite polynomial solutions exist for certain values of $w$ only (None at all for $w=0$ or $w=1$, since eq.(\ref{2m-1_coeff}) cannot vanish for these values).

Of course, one can also solve for $w$ in terms of $m$:
\begin{equation}
w = \frac{m-7 \pm \sqrt{(m-7)^2-4m}}{2 m}    \;\;\;\;\;\;\; {\rm [for} \; m>1 ]    \label{w_from_m}
\end{equation}
or one can write
\begin{equation}
w^2 =  \frac{(m-7)w-1}{m}   \;\;\;\;\;\;\; {\rm [for} \; m>1 ] .   \label{wsq}
\end{equation}

Similarly, for $\tilde{m} < 0$, we can consider lowest power of $r$ in eq.(\ref{MFeq-NS}) and find
\begin{equation}
\tilde{m} = \frac{7w+1}{w(1-w)}      \;\;\;\;\;\;\; {\rm [for} \; \tilde{m}<0 ].         \label{mtilde_lowest}
\end{equation}

A comparison of eqs. (\ref{m_from_w}) and (\ref{mtilde_lowest}) shows that $m>1$ and $\tilde{m}<0$ are not compatible, so that (considering also that $m>\tilde{m}$) we have three possibilities at the top level: (1) $m>1$ and $ \tilde{m} \geq 0$, (2) $\tilde{m}<0$ and $m \leq 1$, (3) $m,\tilde{m} \in \{0,1\}$. The breakdown of the search for all possible finite-polynomial solutions into a complete set of cases is also shown in tabular form in Table \ref{table:NScases}.\\

\begin{table}
 \begin{tabular}{ | l | l | l | l | l }
\hline
\underline{Case 0}: Simple 
& \multicolumn{2}{l}{\underline{Case 0.1}: $w = -1 \rightarrow$}  & $F_{1}(r) = A r^{3} +  C$
  \\ 
 \multicolumn{1}{|r|}{cases giving} & \multicolumn{2}{l}{}  & $F_{2}(r) =  r$
  \\ \cline{2-4} 
 \multicolumn{1}{|r|}{linear equations} & \multicolumn{2}{l}{\underline{Case 0.2}: $w = 0 \rightarrow$}  & $F_{3}(r) = C$
  \\ \hline 
\underline{Case 1}: $m>1$ 
& \underline{Case 1.1}: & \multicolumn{2}{l|}{$w=-1 \rightarrow F(r)=Ar^{3}$: covered by $F_{1}(r)$}  \\  \cline{3-4} 
 \multicolumn{1}{|c|}{$\Rightarrow$} & No $n$  & \multicolumn{2}{l|}{$w=-\frac{1}{3} \rightarrow F_{4}(r)=Ar^{3}$}
  \\  \cline{2-4}  
 \multicolumn{1}{|r|}{$ m = \frac{\textstyle{7w+1}}{\textstyle{w(1-w)}}$,} & \underline{Case 1.2}: & $m = 18$ & $w= 1/2 \rightarrow$ fails (noninteger $n$)
  \\  \cline{4-4} 
 \multicolumn{1}{|r|}{$\tilde{m} \geq 0$} & $n>1 \Rightarrow $ &   & $w= 1/9 \rightarrow$ fails ("brute force")
  \\  \cline{3-4} 
 &  $w$ rational & $m = 15$ & $w= 1/3 \rightarrow$ fails ("brute force")
  \\  \cline{4-4} 
&  &   & $w= 1/5 \rightarrow$ fails ("brute force")
  \\  \cline{3-4} 
 &   & $m = 3$ & $w= -1 \rightarrow$ fails (improper $n$)
  \\  \cline{4-4} 
&  &   & $w= -1/3 \rightarrow$ fails (improper $n$)
  \\  \cline{2-4} 
 & \underline{Case 1.3}: & \multicolumn{2}{l|}{ $B=0 \rightarrow m=3 \rightarrow F(r) = A r^{3} +  C$: same as $F_{1}(r)$}
  \\ \cline{3-4} 
& $n=0$ or 1 &  $B \neq 0$ & $w = -3\pm 2\sqrt{2} \rightarrow$ fails (improper $m$)
  \\ \cline{4-4} 
& $\Rightarrow \tilde{A}=0$ &  & $w=-1/3 \rightarrow F_{5}(r)=Ar^{3} + \frac{3}{2}r$
  \\ \hline 
\underline{Case 2}:  
&  \multicolumn{3}{l|}{$A=0 \rightarrow F_{6}(r) = C$, $w$ arbitrary; covers $F_{3}(r)$}  
  \\ \cline{2-4} 
 \multicolumn{1}{|c|}{$m,\tilde{m} \in \{0,1\}$}
& \multicolumn{3}{l|}{$B=0 \rightarrow F_{7}(r) = \frac{\textstyle{4w}}{\textstyle{w^{2}+6w+1}} r$, $w$ arbitrary, except $-3\pm 2\sqrt{2}$}  
\\  \cline{2-4} 
&  \multicolumn{3}{l|}{$w=-1/5 \rightarrow F_{8}(r)= 5r+B$}
  \\  \hline  
\underline{Case 3}: $\tilde{m}<0 $  &   \multicolumn{3}{|l|}{\underline{Case 3.1}: No $\tilde{n} \rightarrow \tilde{m}=3 \rightarrow$ fails ("brute force") }
  \\  \cline{2-4} 
 \multicolumn{1}{|c|}{$\Rightarrow  m \leq1,$} &   \multicolumn{3}{|l|}{\underline{Case 3.2}: $\tilde{n} < 1 \rightarrow $  fails (rational $w \rightarrow $ positive $\tilde{m}$)}
  \\  \cline{2-4} 
 \multicolumn{1}{|r|}{$ \tilde{m} = \frac{\textstyle{7w+1}}{\textstyle{w(1-w)}}$}&   \multicolumn{3}{|l|}{\underline{Case 3.3}:  $\tilde{n} < 1 \rightarrow $  fails (rational $w \rightarrow $ positive $\tilde{m}$)}
  \\  \hline
\end{tabular}
\caption{Breakdown of all finite polynomial solutions of eq.(\ref{MFeq-NS}) into cases and subcases} 
\label{table:NScases}
 \end{table}

\noindent \underline{Case 1. $m>1 \Rightarrow \;  w \neq 0, \; w \neq 1,  \tilde{m} \geq 0$}. \\

We ask if $n$ exists and if so, in what range its value is.\\

\noindent \underline{Case 1.1. $n$ does not exist  $ \Rightarrow m=\tilde{m} > 1$. }\\

This subcase is amenable to the "brute force" approach, since $F(r)$ has a single term. We get $m=3$, (since $w \neq 0$ and $m \neq 0$), which in turn gives  $w=-1$ or $w = -\frac{1}{3}$. The first solution is the $C=0$ special case of solution 1,  whereas the second one is
\begin{equation}
{\rm {\bf Solution \; 4:}}  \;\;  w =-\frac{1}{3}, \;\; F(r) = A r^{3}.     \label{Sol4} 
\end{equation}
\noindent  \underline{ Case 1.2. $n$ exists,  $ n  > 1$. }\\

In this case, the second-highest power of $r$ in eq.(\ref{MFeq-NS}) is $m+n-1$,  with coefficient
\begin{equation}
(w+1)[m (wn+1)+ n (wm+1)] A B - 2w [m(m-3) + n(n-3)] A B = 0.   \label{m+n-1_coeff}
\end{equation}
giving arbitrary $B$ and, after elimination of  $w^{2}$ by using (\ref{wsq}), the equation
\begin{equation}
(m-n)[(2m-2n-7)w-1] =0  \label{n_from_m}
\end{equation}
which not only gives $n$ in terms of $m$ and $w$, but also means that $w$ is rational.

    A careful inspection of (\ref{w_from_m}) shows that there are only three values of $m$ giving two rational $w$: 18, 15 and 3, with two attendant $w$ values each. The "brute force" approach is applicable now and shows that  all the $m=18$  and $m=15$ cases fail (The $m=3$ cases turn out to belong to case 1.3).\\
    
\noindent  \underline{Case 1.3. $n$ exists, $n=1$ or  $n=0$. }\\

We put $F(r) = A r^{m} + B r + \tilde{A}$ into eq.(\ref{MFeq-NS}), and we use eq.(\ref{m_from_w}) in the coefficients of powers of $r$. Then, the coefficient of $r^{m-1}$ gives
\begin{equation}
 A \tilde{A} m [1+(7-2m)w]  = 0.
\end{equation}
$A$ and $m$ cannot be zero here. $w=1/(2m-7)$, combined with  eq.(\ref{m_from_w}) gives $m=0$ or $m=7/2$, both of which are unacceptable. Therefore, so we must consider $ \tilde{A}=0$. This also makes the constant term in eq.(\ref{MFeq-NS}), $B  \tilde{A} (1+5w)$ vanish. Next we consider the coefficient of $r$,
\begin{equation}
B(-4w+B(w^{2}+6w+1)) = 0,
\end{equation}
which leads to two subcases:\\

\noindent  \underline{Case 1.3.1. $B=0$. }\\

The coefficient of $r^{m}$ reduces to
\begin{equation}
2 A (m-3) m w = 0
\end{equation}
which gives $m=3$, leading to Solution 1.\\

\noindent  \underline{Case 1.3.2. $B=4w/(1+6w+w^{2})$. }\\

Also using eq.(\ref{m_from_w}),
the coefficient of $r^{m}$ this time becomes
\begin{equation}
\frac{2 A (1+3w)(1+6w+w^{2})}
         { (w-1)^{2} w } = 0
\end{equation}
whose $w = -3\pm 2\sqrt{2}$ solutions give $m=1$, therefore are not acceptable for Case 1, whereas the  $w=-1/3$ solution gives
\begin{equation}
{\rm {\bf Solution \; 5:}} \;\;\;\; w=-\frac{1}{3}, \;\; F(r) = A r^3  +  \frac{3}{2}r,      \label{Sol5}
\end{equation}
which does {\em not} include Solution 4 as a special case. 

This finishes Case 1, $m>1$.\\

\noindent  \underline{Case 2. $m,\tilde{m} \in \{0,1\}$. }\\

In this case, $F(r) = A r + \tilde{A}$. Then, the "brute force" approach gives
\begin{equation}
{\rm {\bf Solution \; 6:}} \;\;\;\; w {\rm \; arbitrary}, \;\; F(r) = C= {\rm constant},      \label{Sol6}
\end{equation}
a solution that includes Solution 3;
\begin{equation}
{\rm {\bf Solution \; 7:}} \;\;\;\; w {\rm \; arbitrary (except } -3\pm 2\sqrt{2}\;), \;\; F(r) = \frac{4w}{w^{2}+6w+1}r      \label{Sol7}
\end{equation}
(for $w=-3\pm 2\sqrt{2}$, the left-hand-side of eq.(\ref{MFeq-NS}) cannot vanish at all with $F(r) = Ar$); and 
\begin{equation}
{\rm {\bf Solution \; 8:}} \;\;\;\; w=-\frac{1}{5}, \;\; F(r) = 5 r  +  B.    \label{Sol8}
\end{equation}

\noindent \underline{Case 3. $\tilde{m}<0 \Rightarrow \;  w \neq 0, \; w \neq 1,  m \leq 1$}. \\

Similar to Case 1,\\

\noindent \underline{Case 3.1. $\tilde{n}$ does not exist  $ \Rightarrow m=\tilde{m}$. }\\

Since the "brute force" approach gives $m=3$, there is no solution in this case.\\

\noindent \underline{Case 3.2. $\tilde{n}$ exists, $\tilde{n}<1$ }\\

The coefficient of $r^{\tilde{m}+\tilde{n}-1}$ is given by the same expression as  eq.(\ref{m+n-1_coeff}) with $m \rightarrow \tilde{m}$, $n \rightarrow \tilde{n}$, $A \rightarrow \tilde{A}$ and $B \rightarrow \tilde{B}$. This makes again $w$ rational, but now $\tilde{m}$ must be 18 or 15 or 3, unacceptable because they are positive.\\

\noindent \underline{Case 3.3. $\tilde{n}$ exists, $\tilde{n}=1=m$}\\

The "brute force" approach, with $F(r) = Ar+\tilde{A} r^{\tilde{m}}$ gives the unacceptable (positive) $\tilde{m}$ values 1 and 3.\\

This completes all finite polynomial solutions of equation (\ref{MFeq-NS}). Since Solution 3 is a special case of Solution 7, we will not consider it separately in the following section.

\section{Discussion of the solutions found from the standard (NS) OV equation} \label{sect:NSdisc}

To finalize the solutions, we calculate the metric functions $A(r)$ and $B(r)$ by using (\ref{EEns4}), (\ref{EEns5}), (\ref{eos}) and (\ref{FDefNS}). The calculation of $B(r)$ involves an arbitrary multiplicative constant at the last stage, the change of which is usually interpreted as a rescaling of $t$, therefore physically irrelevant. But such rescaling cannot change the sign of that constant, so we consider the two choices of sign as two separate solutions, unless the requirement of correct signature forces a choice upon us. This happens for solutions 1, 5 and 6, whereas for solutions 4, 7 and 8 we have consider both signs. The results are shown in Table \ref{table:NSsols}, where the well-known solutions are indicated in italics.

\begin{table}
 \begin{tabular}{ | c | p{20 mm}  | c | c | c | p{45 mm} |} \hline
{\bf No.} & \multicolumn{1}{c|}{{\bf $w$}} & {\bf $F(r)$} & {\bf $A(r)=g_{rr}$} & {\bf $B(r)=-g_{tt}$}   & {\bf Comments} \\  \hline \hline 

1 & \multicolumn{1}{c|}{-1} & $Ar^{3}+C$ & \Large $\frac{1}{1-\frac{C}{r}-Ar^{2}}$  & $1-\frac{C}{r}-Ar^{2}$ & {\em K\"{o}ttler (SdS)} \\  \hline

2 & \multicolumn{1}{c|}{-1} & $r$ & $\infty$ & ? & \multicolumn{1}{c|}{--} \\  \hline

4 & \multicolumn{1}{c|}{\large $-\frac{1}{3}$} & $Ar^{3}$ & \Large $\frac{1}{1-Ar^{2}}$ & $\pm 1$  & \mbox{4a+ ($A>0$): {\em ESU}}; \mbox{4a- ($A<0$): open, static}; \mbox{4b: type TD}   \\  \hline

5 & \multicolumn{1}{c|}{$-\frac{1}{3}$} & $Ar^{3}+\frac{3}{2}r$ & \Large $-\frac{1}{\frac{1}{2}+Ar^{2}}$ & \Large $ -\frac{\frac{1}{2}+Ar^{2}}{r^{2}}$  & \mbox{5+ ($A>0$): type TD}; \mbox{5- ($A<0$): BH-like}\\  \hline

6 & arbitrary & $A$ & \Large $\frac{1}{1-\frac{A}{r}}$  &  \large $(1-  \frac{A}{r})$  & {\em Schwarzschild} \\  \hline

7 & arbitrary, except  $-1$, $-3\pm2\sqrt{2}$ & \large $\frac{4w}{w^{2}+6w+1}r$  & \large $\frac{w^{2}+6w+1}{(w+1)^{2}}$  & $\pm$ \Large $(\frac{r}{r_{0}})^{\frac{4w}{w+1}}$ & \mbox{7a: type NS,}  \mbox{incl. static phantom}; \mbox{7b: type TD}\\  \hline

8 & \multicolumn{1}{c|}{$-\frac{1}{5}$} & $5r+B$ & \Large $\frac{1}{\frac{C}{r}-4}$  & \large $\pm \frac{r_{0}}{r}$ & \mbox{8a: type NS};  \mbox{8b: type TD}\\  \hline 
\end{tabular}
\caption{All finite-polynomial solutions of the equation (\ref{MFeq-NS}) for the mass function in the standard (NS) OV case,  together with the corresponding metric functions (Solution 3 does not appear because it is a special case of Solution 7). The well-known solutions are indicated in {\em italics}.
Although we started with the NS OV equation, some of the solutions belong to class TD, as defined in~\cite{revisited}. In TD solutions, $w$ should be replaced by $-w/(1+2w)$. In Solutions 4, 7 and 8, the upper signs in $B(r)$ apply to solutions a and lower signs to solutions b. Note that there is also a Solution~9, coming from the TD OV equation, of type TD or type NS (Sect.~4.1 and sect.~5.1).} 
\label{table:NSsols}
\end{table}

When the metric functions are negative, the spacetime cannot be supported by normal perfect fluid, the source fluid must be tachyonic. In other words, such a spacetime is of type TD in the terminology of \cite{revisited}. In that case, the OV equation, (\ref{ov}), is not valid, but still, $A(r)$-$B(r)$ pairs satisfy the same equation of pressure isotropy for cases NS and TD. Therefore negative metric functions found from NS-equations represent a valid TD solution, but not with the equation of state that one has started with. If the NS equation of state is (\ref{eos}), the corresponding TD equation of state becomes $p=-\frac{w}{1+2w}\rho$.

Solution 1 is the well-known K\"{o}ttler (aka Schwarzschild-de Sitter) solution, the de Sitter part sometimes being called anti-de Sitter if $A$ is negative.

Although Solution 2 satisfies eq.(\ref{MFeq-NS}), it does not correspond to a spacetime: The function $A(r)$ is singular, $B(r)$ is indeterminate\footnote{In fact, if one expresses $F(r)$ as the result of some limit process, e.g. $F(r)=(1+\epsilon)r$ or $F(r)=r+\epsilon$, and takes the limit $\epsilon \rightarrow 0$, the function $B(r)$ and the kind of divergence of $A(r)$ depend on the process used.}.

Solution 4a+ is also well-known: it is the Einstein static universe, with intimate historical connection to the cosmological constant $\Lambda$, equivalent to $w=-1$. But this universe also contains matter ($w=0$), whose attraction is precisely balanced by the repulsion of $\Lambda$. So the matter density is proportional to $\Lambda$ and the net effect is equivalent to a single fluid with $w=-\frac{1}{3}$. Of course, ``in the universe" $Ar^{2}<1$, so $A(r)$ is positive and the signature correct.
Solution 4a- represents an open static universe, albeit with negative energy density, and no coordinate restriction.

The third well-known solution in the table  is Solution 6, Schwarzschild solution. It may at first seem surprising that there is no restriction on $w$. But since $\rho$ vanishes, the value of $w$ does not matter. In other words, it corresponds to a situation where all the fluid --whatever its equation of state parameter is-- has already collapsed to the origin. Also, here we do not apply the usual restriction that $A$ must be positive. If $A$ is negative, the spacetime will give a naked singularity.

Now we turn to the discussion of less well-known solutions in Table \ref{table:NSsols}.\\

\noindent {\em Solution 4b} : \\

$A$ must be positive for correct signature in this solution. It is a dynamic spacetime, $r$ being timelike, (it is solution TD1 of \cite{revisited}) and describes a spacetime that first contracts, then expands in angular directions, while distances in the orthogonal spacelike direction stay fixed\footnote{The KS-like form of the metric is $ds^{2}=-d\tau^{2}+d\rho^{2}+\frac{1}{A} \cosh^{2}(\sqrt{A}\tau) \, d\Omega^{2}$.}. Even though we found this solution for $w=-\frac{1}{3}$, the equation of state is actually $p=\rho$.\\

\noindent {\em Solution 5+} : \\

This solution is also of type TD, contracting in the angular directions and expanding in the orthogonal spacelike direction\footnote{The KS-like form of the metric is $ds^{2}=-d\tau^{2} + A \coth^{2}(\sqrt{A}\tau) d\rho^{2}+\frac{1}{2A} \sinh^{2}(\sqrt{A}\tau) \, d\Omega^{2}$.} as $r \rightarrow 0$, the density also diverges like $\frac{1}{r^{2}}$. The solution can be identified with the $a=1$, $b=-1$, $m=0$, $\frac{1}{R^{2}}=A$ (and the trivial $B=1$ or const=1) choice of Tolman VIII~\cite{Tolman}.\\

\noindent {\em Solution 5-} : \\

Both metric functions switch sign at $r=r_{H}=\sqrt{-\frac{1}{2A}}$, so that the spacetime is static (NS) for $r>r_{H}$ and dynamic (TD)  for $r<r_{H}$. As far as test particle motion is concerned, this spacetime would be that of a black hole; but it must be supported by normal matter in the NS region, and tachyonic matter (with $p=\rho$) in the TD region. As unreasonable as this may seem, it is the only possible perfect fluid interpretation \cite{revisited}. Again, the density diverges like $\frac{1}{r^{2}}$, near the origin (which is in the TD region).  \\

\noindent {\em Solution 7a} : \\

This solution has correct signature only for $A(r)>0$, which means  that the solution is valid except for $-3-2\sqrt{2} < w < -3+2\sqrt{2}$ (and it is of type NS). The cases  $w < -3-2\sqrt{2}$, for example, $w=-6$, represent static (ultra)phantom solutions. The $w=\frac{1}{3}$ case is well-known~\cite[Prob. 23.10]{mtw}; the $w \rightarrow \infty$ limit, meaning zero density but nonzero pressure, is the metric called S1 in \cite{s1}; other valid cases with integer power of $r$ in $B(r)$ are $w=1$ and $w=3$. 

The density is proportional to $\frac{1}{r^{2}}$, but this
 is a mild singularity because the mass function goes to zero as $r \rightarrow 0$, i.e. there is no mass point at the origin. Of course, there is no event horizon, so the singularity is naked.

For test particles, the sign of attraction to the origin is the same as that of $w(1+w)$, that is, the origin
attractive for $w < -3-2\sqrt{2}$ and for $w > 0$, repulsive for $-3+2\sqrt{2} < w < 0$. On the other hand,  the pressure is positive for all $w$ values, and since $p \propto \rho  \propto \frac{1}{r^{2}}$, the pressure gradient is always towards the origin. Naively thinking  in terms of $\rho$, it would seem that both forces would be pushing a fluid element towards the origin in the ultraphantom case ($\rho$ is negative), but the "density of inertial mass" (e.g.~\cite{groundup}), $(\rho + p)$ [here proportional to $w(1+w)$] is positive so that static equilibrium is possible.

These static ultraphantom solutions constitute a counterexample to the impression in the literature (e.g. see~\cite{no_stat_phantom}) that everywhere-phantom static  spherically symmetric solutions cannot exist. 

This solution can be identified with the $n=\frac{2w}{w+1}$, $R \rightarrow \infty$ and $B=r_{0}^{-\frac{2w}{w+1}}$ (or const=$r_{0}^{-\frac{4w}{w+1}}$) choice of Tolman V~\cite{Tolman}.\\

\noindent {\em Solution 7b} : \\

This is a TD solution (a subcase\footnote{Which subcase it is depends on the sign of $w+1$.} of TD2 of \cite{revisited}) valid for $-3-2\sqrt{2} < w < -3+2\sqrt{2}$, except $w=-1$. Assuming $r$ is future-directed, this spacetime expands in the angular directions, and either expands (for $w < -1$) or contracts (for $w > -1$) in the orthogonal spacelike direction\footnote{Metric in KS-like form: $ds^{2}=-d\tau^{2}+\left(\frac{\tau^{2}}{|A|r_{0}^{2}} \right)^{\frac{2w}{w+1}} d\rho^{2} + \frac{\tau^{2}}{|A|} \, d\Omega^{2}$, where \mbox{$A=\frac{w^{2}+6w+1}{(w+1)^{2}}$}.}. An infinite number of $w$-values, crowding -1, exist that give integer power of $r$ in $B(r)$. The density is again proportional to $\frac{1}{r^{2}}$;  the solution can be identified with the almost same subcase of Tolman V~\cite{Tolman} as solution 7a, except\footnote{\label{fn:tolman} Tolman chooses const=$B^{2}$ and later literature reports this form (e.g. \cite{delgaty&lake}).} const=$-r_{0}^{-\frac{4w}{w+1}}$.\\

\noindent {\em Solution 8a} : \\

This solution is type NS. $C$ must be positive and $r<\frac{C}{4}$. Interestingly, radially moving free particles oscillate between a minimum radius and $\frac{C}{4}$ , which may be understood in terms of the repulsion of the negative mass point at the origin ($C=-B$ and $F(r)$ is the mass function) versus the attraction of the fluid, whose ``enclosed active gravitational mass" (e.g.~\cite{groundup}) grows with $r$ (here, both $\rho$ and $\rho + 3p$ are positive).

The origin is a naked singularity, and not only due to the negative point mass there: The scalar curvature  is $\frac{8}{r^{2}}$, that is, it diverges without containing $C$. But, after all, the scalar curvature does not contain $M$ in the Schwarzschild case, either (in fact, it vanishes). $r=\frac{C}{4}$ is a type of boundary, it is a turning point for all radial timelike geodesics.

This solution can be identified with the $n=-\frac{1}{2}$, $R \rightarrow -C$ and $B^{2}=r_{0}$ choice of Tolman V~\cite{Tolman}.\\

\noindent {\em Solution 8b} : \\

This TD solution (with $p=\rho/3$) can be identified with the $n=-\frac{1}{2}$, $R \rightarrow -C$ and 
 restriction for negative $C$, but $r$ must be larger than $\frac{C}{4}$ for positive $C$. 

In the latter case, again $r=\frac{C}{4}$ is a turning point for timelike radial geodesics, but $r$ is timelike, so this spacetime first contracts in the angular directions while expanding in the orthogonal spacelike direction, then the evolution reverses. 

On the other hand, for negative $C$, the spacetime expands in the angular directions while contracting in the orthogonal spacelike direction\footnote{KS-like form of the metric is $ds^{2}=-d\tau^{2}+\frac{r_{0}}{r(\tau)} d\rho^{2} + r^{2}(\tau) \, d\Omega^{2}$, where $\frac{dr}{d\tau}=\pm \sqrt{4-\frac{C}{r}}$}, assuming $r$ is future-directed.

\section{All finite-polynomial solutions for $F(r)$ from the OV-like equations in the TD, ND(KS) and TS cases\label{FindOthSols}}

\subsection{The TD case}

The TD OV equation~\cite{revisited} is
\begin{equation}
p'    =   \frac{(\kappa p r^{3} + F_{TD})}{2 r (r-F_{TD})} (\rho + p) \label{TDov}
\end{equation}
where
\begin{equation}
F_{TD}(r) = - \kappa \int (\rho+2p) r^{2} dr,          \label{FDefTD}
\end{equation}
and the metric functions are found by
\begin{eqnarray}
A(r) & = & \frac{r}{r-F_{TD}(r)}     \label{EEtd4} \\ 
\frac{B'(r)}{B(r)}   & =  & \frac{\kappa p r^{2} + 1}{r-F_{TD}(r)}  -   \frac{1}{r}.  \label{EEtd5}
\end{eqnarray}

The substitution $\tilde{\rho}=-(\rho + 2 p)$ brings the TD OV equation (\ref{TDov})
into the same form as the regular one (\ref{ov}), in terms of $\tilde{\rho}$ and $p$. When expressed in terms of $F_{TD}(r)$, with equation of state (\ref{eos}), we get equation (\ref{MFeq-NS}), but with the replacement $w \rightarrow - \frac{w}{1+2w}$. Since this is another constant equation of state parameter, we will not get any finite-polynomial solutions that are not already in Table \ref{table:NSsols}, {\em unless} $1+2w=0$. This should not be taken as an indication that the TD and TS solutions are trivial relabelings; for more complicated equations of state than eq.(\ref{eos}), the solutions' mathematical forms {\em will} be different. 

For $w=-1/2$, $F_{TD}(r)$ becomes a constant, 
\begin{equation}
{\rm {\bf Solution \; 9:}} \;\;\;\; w=-\frac{1}{2}, \;\; F_{TD}(r) = C.      \label{Sol9}
\end{equation}

\subsection{The ND case}

The ND OV equation~\cite{revisited} is
\begin{equation}
\rho'    =   \frac{3F_{ND}-4r+ \kappa \rho r^{3}}{2 r (r-F_{ND})} (\rho + p) \label{NDov}
\end{equation}
where
\begin{equation}
F_{ND}(r) = - \kappa \int p r^{2} dr          \label{FDefND}
\end{equation}
and the metric functions can be found by
\begin{eqnarray}
A(r) & = & \frac{r}{r-F_{ND}(r)}     \label{EEnd4} \\ 
\frac{B'(r)}{B(r)}   & =  & \frac{1-\kappa \rho r^{2}}{r-F_{ND}(r)}  -   \frac{1}{r}   \label{EEnd5}
\end{eqnarray}

In this case, $F_{ND}(r)$ obeys
\begin{equation}
(1+w) F_{ND}' (3wF_{ND}-4wr-rF_{ND}')+2 w (rF_{ND}''-2F_{ND}')(F_{ND}-r) = 0     \label{MFeq-ND}
\end{equation}

To find all finite-polynomial solutions of this equation (dropping label {\em ND}), we follow the same procedure as in Section \ref{FindStdSols}, and show the results in Table \ref{table:NDcases}.

\begin{table}
 \begin{tabular}{ | l | l | l | l | l }
\hline
\underline{Case 0}: Simple 
& \multicolumn{2}{l}{\underline{Case 0.1}: $w = -1 \rightarrow$}  & $F_{10}(r) = A r^{3} +  C$
  \\ 
 \multicolumn{1}{|r|}{cases giving} & \multicolumn{2}{l}{}  & $F_{11}(r) =  r$
  \\ \cline{2-4} 
 \multicolumn{1}{|r|}{linear equations} & \multicolumn{2}{l}{\underline{Case 0.2}: $w = 0 \rightarrow$}  & $F_{12}(r) = C$
  \\ \hline 
\underline{Case 1}: $m>1$ & \underline{Case 1.1}: & \multicolumn{2}{l|}{\underline{C. 1.1.1}: No $n \rightarrow$ fails ("brute force")} \\  \cline{3-4} 

 \multicolumn{1}{|c|}{$\Rightarrow$} & $w=1$  & \multicolumn{2}{l|}{\underline{C. 1.1.2}: $n>1 \rightarrow$ fails (coefficient of $r^{m+n-1}$)}
   \\  \cline{3-4}  
   
$(w-1) \cdot $ &  &   \multicolumn{2}{l|}{\underline{C. 1.1.3}: $n=1$ or 0 $\rightarrow$ fails ("brute force")}
  \\  \cline{2-4} 
  
 \multicolumn{1}{|r|}{$(3w+m)=0$,} & \underline{Case 1.2}: & \multicolumn{2}{l|}{\underline{C. 1.2.1}: No $n \rightarrow w=-1, F(r)=Ar^{3}$ (covered by $F_{10}$)}
  \\  \cline{3-4} 
  
 \multicolumn{1}{|r|}{$\tilde{m} \geq 0$} & $w=-m/3$ & \underline{C. 1.2.2}: $n>1$ & No $p \rightarrow$ fails ("brute force")
  \\  \cline{4-4} 
  
 &   & $\rightarrow m = 2 n +3$ & $p$ exists $\rightarrow$ fails (coefficient of $r^{m+p-1}$)
  \\  \cline{3-4} 
  
&  & \underline{C. 1.2.3}: $n=1$  & \underline{C. 1.2.3.1}: $m=3 \rightarrow B=0 \rightarrow $
  \\ 
  
 &  &   \multicolumn{1}{|r|}{or 0} & \multicolumn{1}{r|}{$w=-1, F(r)=Ar^{3}+C$: same as $F_{10}$}
  \\ \cline{4-4} 
  
&  &  & \underline{C. 1.2.3.2}: $C=0 \rightarrow$ 
  \\ 
&  &  & 1) $m=3 \rightarrow B=0$: covered by $F_{10}$
  \\ 
 &  &   & 2) $w=-3, F_{13}(r)=Ar^{9}+\frac{\textstyle{9}}{\textstyle{8}}r$
  \\ \hline
  
\underline{Case 2}:  
&  \multicolumn{3}{l|}{$w$ arbitrary, $F_{14}(r) = C$, ; covers $F_{12}$.}  
  \\ \cline{2-4} 
  
 \multicolumn{1}{|c|}{$m,\tilde{m} \in \{0,1\}$}
& \multicolumn{3}{l|}{$F_{15}(r) = \frac{\textstyle{4w^{2}}}{\textstyle{3w^{2}-2w-1}} r$, $w$ arbitrary, except $-\frac{1}{3}$ and 1}  
\\  \cline{2-4} 

&  \multicolumn{3}{l|}{$w=\frac{1}{3} \rightarrow F_{16}(r)= -\frac{1}{3}r+B$}
  \\  \hline  
  
\underline{Case 3}: $\tilde{m}<0 $  &   \underline{Case 3.1}: & \multicolumn{2}{l|}{\underline{C. 3.1.1}: No $\tilde{n} \rightarrow F_{17}(r)= \frac{\textstyle{C}}{\textstyle{r}}, w=1$} \\  \cline{3-4} 

 \multicolumn{1}{|c|}{$\Rightarrow$} &   $w=1$  & \multicolumn{2}{l|}{\underline{C. 3.1.2}: $\tilde{n}<1 \rightarrow$ fails (coefficient of $r^{m+n-1}$)}
   \\  \cline{3-4}  

$(w-1) \cdot $ &     &   \multicolumn{2}{l|}{\underline{C. 3.1.3}: $\tilde{n}=1$  $\rightarrow$ fails ("brute force")}
  \\  \cline{2-4} 

 \multicolumn{1}{|r|}{$(3w+\tilde{m})=0$,} & \underline{Case 3.2}: & \multicolumn{2}{l|}{\underline{C. 3.2.1}: No $\tilde{n} \rightarrow  \tilde{m}=3 \rightarrow$ fails}
  \\  \cline{3-4} 

 \multicolumn{1}{|r|}{$m \leq 1$} &  $w=-\tilde{m}/3$ & \multicolumn{2}{l|}{\underline{C. 3.2.2}: $\tilde{n}=1$   $\rightarrow$ fails ("brute force", $\tilde{m}$=1,9)}
   \\  \cline{3-4} 

&  & \underline{C. 3.2.3}: $\tilde{n}<1$  & \underline{C. 3.2.3.1}: No $p \rightarrow$ fails ("brute force")
  \\  \cline{4-4} 
  
 &  &   \multicolumn{1}{|r|}{$\Rightarrow \tilde{m} = 2 \tilde{n} +3$} & \underline{C. 3.2.3.2}: $p<-3 \rightarrow$ fails ($\tilde{p}=\tilde{n}$)
  \\  \cline{4-4} 
  
&  &  (coefficient of & \underline{C. 3.2.3.3}: $p>-3 \rightarrow$ fails ($\tilde{p}=\tilde{m}$)
  \\  \cline{4-4} 

&  &   \multicolumn{1}{|r|}{$r^{\tilde{m}+\tilde{n}-1}$)} & \underline{C. 3.2.3.4}: $p=-3 \rightarrow$ fails ("brute force")  \\  \hline   
\end{tabular}
\caption{Breakdown of all finite polynomial solutions of eq.(\ref{MFeq-ND}) into cases and subcases} 
\label{table:NDcases}
 \end{table}

\subsection{The TS case}

The TS OV equation~\cite{revisited} is
\begin{equation}
\rho' + 2p'  =  \frac{3F_{TS}-4r- \kappa (\rho+2p) r^{3}}{2 r (r-F_{TS})} (\rho+p)      \label{TSov}
\end{equation}
where
\begin{equation}
F_{TS}(r) = - \kappa \int p r^{2} dr ,          \label{FDefTS}
\end{equation}
and the metric functions are found by
\begin{eqnarray}
A(r) & = & \frac{r}{r-F_{TS}(r)}     \label{EEts4} \\ 
\frac{B'(r)}{B(r)}   & =  & \frac{1+\kappa (\rho+2p) r^{2}}{r-F_{TS}(r)}  -   \frac{1}{r}   \label{EEts5}
\end{eqnarray}

Again, for $F_{TS}(r)$, with equation of state (\ref{eos}) we get equation (\ref{MFeq-ND}), but with the replacement $w \rightarrow - \frac{w}{1+2w}$, leading to the only potentially new solutions for $1+2w=0$. Then, we get two solutions: 
\begin{eqnarray}
{\rm {\bf Solution \; 18:}} & \;\;\;\; w=-\frac{1}{2}, &  \;\; F_{TS}(r) = C, \\      \label{Sol18}
{\rm {\bf Solution \; 19:}} & \;\;\;\; w=-\frac{1}{2}, &  \;\; F_{TS}(r) = \frac{4}{3} r.      \label{Sol19}
\end{eqnarray}

\section{Discussion of solutions found from the OV-like equations in the other cases} \label{sect:NonStdSols}

In this section,  we calculate the metric functions $A(r)$ and $B(r)$ for each solution from the previous section by using the relevant formulae, and discuss the solutions.

\subsection{The TD case}

For Solution 9, we get
\begin{equation}
A(r)  =  \frac{1}{1-C/r}     \label{Soltd3A}\\ 
\end{equation}
which for $r < C$ (only possible if $C > 0$) gives
\begin{equation}
B(r)   =  - r_{1}^{-4}  \left[ (2 r^{2}+5Cr-15 C^{2})+\sqrt{\frac{C-r}{r}} \left(C_{1}-15 C^{2} \tan^{-1}\sqrt{\frac{C-r}{r}}\right) \right]^{2},    \label{SolTd3B}
\end{equation}
i.e. Solution TD3 of \cite{revisited}.

On the other hand, for $r > C$ we find
\begin{equation}
B   =  r_{1}^{-4} \left[ (2 r^{2}+5Cr-15 C^{2})+\sqrt{\frac{r-C}{r}} \left(C_{1}+15 C^{2}  \ln(\frac{\sqrt{r-C}+\sqrt{r}}{\sqrt{|C|}})\right)\right]^{2} ,    \label{SolNSB}
\end{equation}
the solution called NS1 in \cite{revisited}, found in \cite{Kuch68I} and named Kuch68 I in~\cite{delgaty&lake}. It describes a spacetime where pure pressure is in static equilibrium with its own gravitational attraction.

\subsection{The ND(KS) and TS cases}

The solutions found for the ND(KS) and TS cases, together with their metric functions, are shown in Table \ref{table:NDsols} (Solutions 12 and 18 do not appear because they are special cases of Solution 14). As in Sect.3, the sign of $B(r)$ is arbitrary, unless forced by the signature requirement.

\begin{table}
 \begin{tabular}{ | c | p{17 mm}  | c | c | c | p{36 mm} |} \hline
{\bf No.} & \multicolumn{1}{c|}{{\bf $w$}} & {\bf $F(r)$} & {\bf $A(r)=g_{rr}$} & {\bf $B(r)=-g_{tt}$}   & {\bf Comments} \\  \hline \hline 

10 & \multicolumn{1}{c|}{$-1$} & $Ar^{3}+C$ & $\frac{\textstyle{1}}{\textstyle{1-\frac{\textstyle{C}}{\textstyle{r}}-Ar^{2}}}$ & $1-\frac{\textstyle{C}}{\textstyle{r}}-Ar^{2}$ & {\em K\"{o}ttler (SdS)} \\  \hline

11 & \multicolumn{1}{c|}{-1} & $r$ & $\infty$ & ? & \multicolumn{1}{c|}{--} \\  \hline

13 & \multicolumn{1}{c|}{ $-3$} & $Ar^{9}+\frac{\textstyle{9}}{\textstyle{8}}r$ & $-\frac{\textstyle{8}}{\textstyle{1+8Ar^{8}}}$  & $ -\frac{\textstyle{1+8Ar^{8}}}{\textstyle{r^{6}}}$  & \mbox{13+ ($A>0$):}  \mbox{type ND(KS)}, phantom-filled dynamic universe; \mbox{13- ($A<0$): BH-like}   \\  \hline

14 & arbitrary & $C$ & $\frac{\textstyle{1}}{\textstyle{1-\frac{\textstyle{C}}{\textstyle{r}}}}$  & $1-\frac{\textstyle{C}}{\textstyle{r}}$ & {\em Schwarzschild}\\  \hline 

15 & arbitrary, except $-\frac{1}{3}$ and $1$ & $\frac{\textstyle{4w^{2}}}{\textstyle{3w^{2}-2w-1}}r$ & $-\frac{\textstyle{(w-1)(3w+1)}}{\textstyle{(w+1)^{2}}}$  & $\pm$ \Large $ (\frac{r}{r_{0}})^{-\frac{4w}{w+1}}$  & \mbox{15a ($-\frac{1}{3}<w<1$):} \mbox{type TS};  \mbox{15b (otherwise):} \mbox{type ND(KS)}, incl. DE, incl. phantom\\  \hline

16 & \multicolumn{1}{c|}{$\frac{1}{3}$} & $-\frac{1}{3}r+B$  &  $\frac{\textstyle{3r}}{\textstyle{4r-3B}}$  & \large $\pm \frac{\textstyle{r_{0}}}{\textstyle{r}}$  & \mbox{16a: type TS;} \mbox{16b: type ND(KS)}\\  \hline

17 &  \multicolumn{1}{c|}{$1$} & $\frac{\textstyle{C}}{\textstyle{r}}$ &  $\frac{\textstyle{1}}{\textstyle{1-\frac{\textstyle{C}}{\textstyle{r^{2}}}}}$ & $\pm 1$ &  \mbox{17a: type TS;} \mbox{17b: type ND(KS)}\\  \hline

19 & \multicolumn{1}{c|}{$-\frac{1}{2}$} & $\frac{4}{3}r$ & $-3$ & $ -(\frac{\textstyle{r}}{\textstyle{r_{0}}})^{-4}$  & type ND(KS)\\  \hline
\end{tabular}
\caption{All finite-polynomial solutions for $F(r)$ in the ND(KS) and TS cases as defined in~\cite{revisited};  together with the corresponding metric functions (Solutions 12 and 18 do not appear because they are special cases of Solution 14). In Solutions 15, 16 and 17, the upper signs in $B(r)$ apply to solutions a and lower signs to solutions b. The well-known solutions are indicated in {\em italics}.} 
\label{table:NDsols}
\end{table}

The Schwarzschild and K\"{o}ttler (SdS) solutions, which appeared in Table \ref{table:NSsols}, are found in this table as well, because they cannot really be classified in this scheme. Our classification is based upon the nature and direction of motion of the fluid, but for these solutions, the stress-energy-momentum tensor is independent of the fluid four-velocity: The $u_{\mu}u_{\nu}$ term in $T_{\mu\nu}$ is multiplied by $p+\rho$; and  $p+\rho=0$ for the K\"{o}ttler solution, $p=\rho=0$ for Schwarzschild. Hence,  these solutions satisfy the equations for all four cases.

The other solutions in the table are less well-known: \\

\noindent {\em Solution 13+} : \\

This solution is type ND (KS), representing a dynamic spacetime filled with a phantom perfect fluid. Assuming $r$ is future-directed, the spacetime expands in the angular directions; in the perpendicular spacelike direction, it first contracts, reaches a minimum, then expands\footnote{The KS form of the metric is $ds^{2}=-d\tau^{2}+\frac{1+8Ar^{8}(\tau)}{r^{6}(\tau)} d\rho^{2} + r^{2}(\tau) \, d\Omega^{2}$, where \mbox{$\frac{dr}{d\tau}=\pm \sqrt{Ar^{8}+\frac{1}{8}}$.}}. It is singular at both ends of the evolution, that is, at $r=0$ and as $r \rightarrow \infty$, the first singularity being in the finite past, the second in the infinite future. Of course, these attributes switch if $r$ is past-directed.\\

\noindent {\em Solution 13-} : \\

This solution,  like Solution 5-, represents a black hole spacetime, as far as test particle motion is concerned; but it must be supported by two different fluids on the two sides of the horizon: tachyonic fluid in the outside, static region and normal fluid in the dynamic region inside/in the future.\\

\noindent {\em Solutions 15a} : \\

This  is a TS solution. For positive $w$,  that is, for $0<w<1$  radially incoming test particles are reflected near the origin back to infinity, whereas for negative $w$, that is, $-\frac{1}{3}<w<0$, the origin constitutes a potential well from which they cannot escape.\\

\noindent {\em Solutions 15b} : \\

This solution is identical to the $C_{1}=0$ special case of Solution ND2 of \cite{revisited}. If $r$ is future-directed, it expands in the angular directions, and it expands or contracts in the perpendicular spacelike direction, if the sign of $\frac{w}{w+1}$ is negative or positive, respectively\footnote{The KS form of the metric is $ds^{2}=-d\tau^{2}+\left(\frac{\tau^{2}}{r_{0}^{2}|A|} \right)^{-\frac{2w}{w+1}} d\rho^{2} +\frac{\tau^{2}}{|A|} \, d\Omega^{2}$, where \mbox{$A=-\frac{(w-1)(3w+1)}{(w+1)^{2}}$}.}. Note that this means expansion for non-phantom dark energy ($-1<w<-\frac{1}{3}$) and ``radial'' contraction for phantom energy.\\

\noindent {\em Solution 16a} : \\

This  is a TS solution, where we must have $4r>3B$, i.e. we have a restriction on $r$ if $B$ is positive. Either way, the equation of motion for test particles shows that tachyonic $w=\frac{1}{3}$ fluid is repulsive, consistent with the solution {\em 15a}.\\

\noindent {\em Solution 16b} : \\

This is an ND (KS) solution, where $4r<3B$. It represents a radiation-filled universe that expands and recollapses in angular directions while contracting and reexpanding in the perpendicular spacelike direction\footnote{The KS form of the metric is $ds^{2}=-d\tau^{2}+\frac{r_{0}}{r(\tau)} d\rho^{2} + r^{2}(\tau) \, d\Omega^{2}$, where \mbox{$\frac{dr}{d\tau}=\pm \sqrt{\frac{B}{r}-\frac{4}{3}}$}. The arbitrary $r_{0}$ must be chosen as $\frac{3}{4}B$ for agreement with \cite{kantowskiTH}, p.1684.}; first found in \cite{kantowskiTH}. \\

\noindent {\em Solutions 17a} : \\

This solution is identical to solution TS1 of \cite{revisited}, where we must have $r^{2}>C$. For positive $C$, $r_{0}=\sqrt{C}$ is a turning point for radial geodesics; for negative $C$, there are no such turning points.\\

\noindent {\em Solutions 17b} : \\

This is solution {\em ND1} of \cite{revisited}, apparently first found in \cite{thorneTH}, describing a finite-lifetime universe containing stiff matter,  expanding and recollapsing in the angular directions\footnote{The KS form of the metric is $ds^{2}=-d\tau^{2}+ d\rho^{2} + (C-\tau^{2}) \, d\Omega^{2}$.}.\\

\noindent {\em Solution 19} : \\

This solution is the $C_{1}=0, A=-3$ special case of solution {\em ND2} of \cite{revisited}, describing a spacetime containing pressure, but no density (because it is an ND (KS) solution found from the TS equations, its equation of state is {\em not} $p=-\frac{\rho}{2}$); expanding in angular directions while contracting in the perpendicular spacelike direction\footnote{The KS form of the metric is $ds^{2}=-d\tau^{2}+\left(\frac{\tau_{0}}{\tau}\right)^{4} d\rho^{2} + \frac{\tau^{2}}{3} \, d\Omega^{2}$.}, if $r$ is taken to be future-directed. 

\section{Finite-polynomial $A(r)$?}   \label{sect:PolA}

Another possible way to look for solutions is to work in terms of $A(r)$ rather than $F(r)$ by using equation (\ref{EEns4}). This leads to an equation with terms second to fourth order in $A(r)$ and/or its derivatives. In trying to find a finite-polynomial solution for $A(r)$, if the highest power of $r$ in $A(r)$ is $m$, the highest power of $r$ in the equation is $4m+1$; but it is multiplied by $A^{2} (w+1)^{2}$ in cases NS and TD, and $-A^{2} (w+1)^{2}$ in cases ND and TS; unless $m=0$. Setting the trivial $w=-1$ case aside,  therefore, the highest possible value for $m$ is zero. A similar argument shows that the lowest power in the $A(r)$ polynomial must be zero or higher. Hence the only finite polynomial $A(r)$ can be for equation of state (\ref{eos}) is a constant.

\section{Summary and Conclusions \label{sect:conc}}

We have considered spherically symmetric perfect fluid solutions in General Relativity and
 found {\em all} finite-polynomial solutions -including negative powers- of the equation satisfied by the so-called "mass function" and its mathematical analogs for the equation of state $p=w\rho$; and discussed the associated spacetimes. 
 
The equation for the mass function follows from the Oppenheimer-Volkoff (OV) equation in the standard case where the fluid is static and normal (i.e. timelike fluid four-velocities, $u^{\mu} u_{\mu} = -1$). However, the metric ansatz used in that analysis can also accomodate cases where the spacetime is dynamic in a certain way, or the fluid is tachyonic; as discussed in \cite{revisited}. In these other cases analogous, but different functions exist, satisfying their own equations. 

The solutions we found for the standard case, NS, are mathematically not very original; they are either some limiting cases of solutions found long ago by Tolman \cite{Tolman} or simple modifications thereof. Some aspects of the physical nature of these solutions can be seen in new light however, considering the classification in \cite{revisited} and newly cosmologically relevant concepts of dark energy and phantom energy. 
The solutions (Table \ref{table:NSsols}) include dynamic spacetimes supported by tachyonic fluids ({\em 4b, 5+, 7b} and  {\em 8b}) and a static spacetime containing a $w=-\frac{1}{5}$ fluid around a negative point mass ({\em 8a}). 
The TD case gives two extra solutions, one describing a spacetime where pure pressure is in static equilibrium with its own gravitational attraction.

Some interesting solutions are also found from the ND(KS) and TS cases (Table \ref{table:NDsols}): There are {\em static} solutions supported by {\em tachyonic} fluids ({\em 15a, 16a, 17a}), the first two presumably original.  Some solutions ({\em 13+, 15b, 16b, 17b, 19}) are of the Kantowski-Sachs (KS) class: Solutions {\em 16b, 17b} and {\em 19} describe dynamic KS-universes containing radiation, stiff matter and pure pressure, respectively. 

We would like to particularly point out the following solutions:

\begin{itemize}
\item Solution {\em 5-}  is a black hole-like spacetime, which must be supported by normal matter outside the horizon and tachyonic fluid on the inside.

\item Solution {\em 7a} for $w<-3-2\sqrt{2}$ represents, perhaps unexpectedly, a family of {\em static} ``ultraphantom'' solutions.

\item Solution {\em 13+} is a {\em phantom} KS solution, probably new.

\item Solution {\em 13-} is similar to Solution {\em 5-}, a black hole-like spacetime, supported by segregated normal and tachyonic matter, except in this solution, the tachyonic fluid is outside and normal fluid is inside. It was concluded in \cite{revisited} that black holes supported by perfect fluids cannot be ``simple''.

\item Solutions {\em 13+} and {\em 13-} both have by ``mass functions'' containing the $9^{\rm th}$ power of $r$,  therefore $g_{rr}$ containing $r^{8}$ in the denominator and  $g_{tt}$ containing $r^{-6}$.

\item Solution {\em 15b} can also be valid for dark energy, including phantom, exhibiting anisotropic expansion for non-phantom dark energy.

\end{itemize}

There are {\em no other solutions} where $F(r)$ is a finite polynomial of $r$ for the assumed equation of state. One can also express the problem(s) in terms of $A(r)$, and then try to find finite polynomial solutions. The only such solution is $A(r)$=constant.

\section*{Acknowledgments}

We would like to thank to M. \"{O}zbek and N.M. Uzun for helpful discussions. This work was partially supported by Grant No. 06B303 of the Bo\u{g}azi\c{c}i University Research Fund.


\begin{thebibliography}{99}

\bibitem{mtw} C.W. Misner, K.S. Thorne and J.A. Wheeler, {\it Gravitation},   Freeman, New York (1973).

\bibitem{de}  M.S. Turner, D. Huterer, "Cosmic Acceleration, Dark Energy and Fundamental Physics", {\em Journal of the Physical Society of Japan}, {\bf 76}, 111015  (2007) [{\em arXiv}: 0706.2186].

\bibitem{phantom}  R. R. Caldwell, "A phantom menace? Cosmological consequences of a dark energy component with super-negative equation of state", {\em Physics Letters B}, 
{\bf 545}, 23-29 (2002) [{\em arXiv}: astro-ph/9908168].

\bibitem{acceleration-hiZsst}  A.G. Riess et al. (High-z Supernova Search Team), "Observational evidence from supernovae for an accelerating universe and a cosmological constant", {\em Astronomical Journal} {\bf 116}, 1009-1038 (1998).

\bibitem{acceleration-SCP}  S. Perlmutter et al. (Supernova Cosmology Project), "Measurements of Omega and Lambda from 42 high redshift supernovae", {\em Astrophysical Journal} {\bf 517}, 565-586 (1999).

\bibitem{ov}  J.R. Oppenheimer and G.M. Volkoff, "On Massive Neutron Cores", {\em Physical Review}  {\bf 55}, 374 - 381 (1939).

\bibitem{ks}  R. Kantowski  and R.K. Sachs, "Some Spatially Homogeneous Anisotropic Relativistic Cosmological Models", {\em Journal of Mathematical Physics}, {\bf 7}, 443-446 (1966).

\bibitem{exsols}  H. Stephani, D. Kramer, M. MacCallum, C. Hoensealers, E. Herlt, {\em Exact Solutions of Einstein's Equations}, 2$^{\rm nd}$ Ed, Cambridge U.P. (2003).

\bibitem{revisited}  \.{I}. Semiz, "The standard 'static' spherically symmetric ansatz with perfect fluid source revisited", {\em International  Journal of Modern Physics} {\bf 19}, 1  (2010).

\bibitem{s1}  P. Boonserm, M. Visser, and S. Weinfurtner, "Generating perfect fluid spheres in general relativity", {\em Physical Review D} {\bf 71}, 124037  (2005).

\bibitem{groundup} B. Schutz, {\it Gravity from the ground up},   Cambridge U.P. (2003).

\bibitem{no_stat_phantom} V. Dzhunushaliev, V. Folomeev, R. Myrzakulov, D. Singleton, "Non-singular solutions to Einstein-Klein-Gordon equations with a phantom scalar field", {\em Journal of High Energy Physics}, {\bf 07}, 094  (2008) [{\em arXiv}: 0805.3211].

\bibitem{Tolman}  R.C. Tolman, "Static Solutions of Einstein's Field Equations for Spheres of Fluid", {\em Physical Review} {\bf 55}, 364 - 373 (1939).

\bibitem{delgaty&lake}  M.S.R. Delgaty and K. Lake,  "Physical acceptability of isolated, static, spherically symmetric, perfect fluid solutions of  Einstein's equations", {\em Comp. Phys. Commun.} {\bf 115}, 395-415 (1998); also  {\it arXiv}: gr-qc/9809013. 

\bibitem{Kuch68I}  B. Kuchowicz, "Extensions of external Schwarzschild solution", {\em Bulletin de l'Academie Polonaise des Sciences-Serie des Sciences Mathematiques, Astronomiques et Physiques}, {\bf 16}, 341 (1968).

\bibitem{kantowskiTH} R. Kantowski, "Some Relativistic Cosmological Models" (Dissertation, University of Texas, 1966); reprinted in {\em General Relativity and Gravitation}, {\bf 30}, 1665-1700 (1998).

\bibitem{thorneTH} K.S. Thorne, "Geometrodynamics of Cylindrical Systems", (Dissertation, Princeton University, 1965).


\end{thebibliography}
\end{document}